\documentclass[aps,twocolumn,superscriptaddress,preprintnumbers,nofootinbib]{revtex4-1}
\usepackage{graphicx,amssymb,amsmath,color}
\usepackage{hyperref}
\usepackage{multirow}
\usepackage{color}

\newcommand{\beq}{\begin{equation}}
\newcommand{\eeq}{\end{equation}}
\newcommand{\bea}{\begin{eqnarray}}
\newcommand{\eea}{\end{eqnarray}}
\newcommand{\barr}{\begin{array}}
\newcommand{\earr}{\end{array}}
\newcommand{\bc}{\begin{center}}
\newcommand{\ec}{\end{center}}
\newcommand{\bit}{\begin{itemize}}
\newcommand{\eit}{\end{itemize}}
\newcommand{\ben}{\begin{enumerate}}
\newcommand{\een}{\end{enumerate}}

\newcommand{\nn}{\nonumber}

\newcommand{\br}{{\rm Br}}

\newcommand{\fb}{{\,{\rm fb}}}

\newcommand{\gev}{{\;{\rm GeV}}}
\newcommand{\tev}{{\;{\rm TeV}}}

\newcommand{\sg}{\sigma}

\newcommand{\lm}{\lambda}

\newcommand{\tb}{\tan\beta}

\newcommand{\cba}{\cos(\beta-\alpha)}
\newcommand{\sba}{\sin(\beta-\alpha)}
\newcommand{\yh}{\hat{y}}

\newcommand{\sghat}{{\hat \sigma}}
\newcommand{\shat}{{\hat s}}

\newcommand{\rr}{{\gamma \gamma}}
\newcommand{\ttop}{{t\bar{t}}}
\newcommand{\ttbar}{t{\bar t}}
\newcommand{\bb}{{b\bar{b}}}
\newcommand{\ttau}{{\tau^+ \tau^-}}

\newcommand{\Fig}[1]{Fig.~\ref{#1}}
\newcommand{\Eq}[1]{Eq.~(\ref{#1})}
\newcommand{\Sec}[1]{Sec.~\ref{#1}}

\newcommand{\calA}{{\cal A}}

\def\GeV{\,{\rm GeV}}

\begin{document}

\title{
Dip or nothingness of a Higgs resonance \\from the  interference with a complex phase
}

\author{Sunghoon Jung}
\email{nejsh21@gmail.com}
\affiliation{School of Physics, Korea Institute for Advanced Study, Seoul 130-722, Korea}

\author{Jeonghyeon Song}
\email{jeonghyeon.song@gmail.com}
\affiliation{School of Physics, KonKuk University, Seoul 143-701, Korea}

\author{Yeo Woong Yoon}
\email{ywyoon@kias.re.kr}
\affiliation{School of Physics, KonKuk University, Seoul 143-701, Korea}

\begin{abstract}

We show that new resonance shapes -- a pure dip, nothingness and an enhanced pure peak -- can be produced from the interference between resonance and continuum with a relative phase. Production conditions of those new shapes are derived based on a general parameterization of the interference. The narrow width approximation is modified to work with the non-zero imaginary part of interference, and the correction factor can characterize the resonance shape. We demonstrate that the new resonance shapes of heavy Higgs bosons, $H^0$ and $A^0$ in the Type II aligned two Higgs doublet model, generally show up in $gg \to H^0/A^0 \to \ttop$ as well as $\bb$ and $\rr$ channels. The pure $A^0$ resonance dip in the $\ttop$ channel is a particularly interesting signal as it can be probed well by the current search techniques that do not even take into account interferences; the high-luminosity LHC 14 TeV can perhaps probe a large part of its parameter space.

\end{abstract}

\preprint{KIAS-P15020}

\maketitle

\section{Introduction}

Many exciting discoveries in particle physics history
came with \emph{excesses} above continuum backgrounds.
The existence of a new resonance such as $J/\psi$ mesons~\cite{Aubert:1974js,Augustin:1974xw},
$W$ bosons~\cite{Arnison:1983rp,Banner:1983jy},
$Z$ bosons~\cite{Arnison:1983mk,Bagnaia:1983zx},
top quarks~\cite{Abe:1995hr,Abachi:1995iq},
and the recently discovered Higgs boson~\cite{Aad:2012tfa,Chatrchyan:2012ufa},
was all confirmed by clear excesses
or a resonance peak in the invariant or transverse mass distribution.
Most new physics searches at the Large Hadron Collider (LHC)
are also focused primarily on excesses.

A pure Breit-Wigner (BW) resonance peak is, however, modified from the interference
with a continuum or other resonance processes.
If the interference is purely real
(except for the small imaginary part in a BW propagator),
a symmetric dip-peak or peak-dip structure appears and adds to a BW resonance peak.
As for the Standard Model (SM) Higgs boson in the $\rr$ channel at the LHC,
the real-part interference
shifts the location of the resonance peak, affecting the pole mass measurements~\cite{Martin:2012xc}. But a resonance peak never disappears from the purely real interference.

The modification on the pure BW resonance peak can be much more significant
and come with a more variety of shapes if the interference involves a significant imaginary part.
The dip-like resonance shape induced from the imaginary part in elementary particle physics
was first theoretically glimpsed
for the heavy Higgs boson in $gg \to H \to \ttop$~\cite{Gaemers:1984sj}.
The complex phase arises from the loop of top quarks as the Higgs boson is heavier than the top pair threshold.
More dedicated study in this channel~\cite{Dicus:1994bm} found
that various dip-like resonances
can appear for a wide range of the heavy Higgs boson mass in
the two Higgs doublet model (2HDM) and
the minimal supersymmetric standard model (MSSM).
Since then, the resonance-continuum interferences including imaginary parts have been
calculated in photon collider processes:
$\rr \to H \to WW$~\cite{Morris:1993bx}, $ZZ$~\cite{Niezurawski:2002jx},
$t\bar{t}$~\cite{Asakawa:1999gz} and $b\bar{b}$~\cite{Dixon:2008xc} (as well as $\ttop \to ZZ$~\cite{Basdevant:1992nb}).
A variety of resonance shapes were found to appear, composed of dips and peaks with various depths and heights.
On the other hand,
hadron collider studies for $gg \to H \to \rr$~\cite{Dicus:1987fk,Martin:2012xc,Dixon:2003yb},
$WW$~\cite{Campbell:2011cu,Kauer:2015hia} and $ZZ$~\cite{Glover:1988rg,Kauer:2015hia}
found only peak-like shapes of the SM-like heavy Higgs boson.
The CP violation, another source of a relative phase, was also considered in the resonance-continuum interference~\cite{Bernreuther:1998qv}, but the studies were not focused on the resonance dip.
Meanwhile, dip-like structures were observed in hadronic physics (see references in Ref.\cite{Dicus:1994bm}) and were also expected to be produced from time-like form factors in some particle physics models~\cite{Bai:2014fkl}.

The diversity of resonance shapes due to the imaginary part of interference
complicates new resonance searches. In particular, the usual narrow width approximation (NWA) does not apply with non-zero imaginary parts, further hindering collider studies. Perhaps, even not all possible resonance shapes have been systematically identified and studied. Although resonance dip structures were analyzed by $J=0$ partial wave amplitudes for certain processes~\cite{Dicus:1994bm,Chiu:1994se}, a generalization to other processes is not straightforward without dedicated calculations. 

In this paper, we first aim to provide a general description of the resonance-continuum interference with a relative phase in \Sec{sec:general}. Three most striking new resonance shapes are identified -- pure dips without associated peaks, nothingness and enhanced pure peaks -- and their mathematical conditions are derived.  Based on the description, we find it possible to modify the NWA by introducing a multiplicative correction factor $C$, which can also be used to tell shape characteristics. They will be defined in \Sec{sec:general}, and will be used to demonstrate and discuss various new resonance shapes of heavy Higgs bosons, $H^0$ and $A^0$ in
the CP-conserving Type II 2HDM,
in $\ttop, \, \bb$ and $\rr$ channels (\Sec{sec:2HDM:overview}). In \Sec{sec:higgschannels}, various other Higgs processes will also be briefly discussed in regard of whether pure dips can be produced.

The pure resonance dip is a particularly interesting signal. As the pure dip has the BW shape (with just a negative sign), its signal is well localized and searches are eased; most importantly, it can be probed by the current searches that do not even take into account interference effects. In \Sec{subsec:tt}, we will discuss LHC prospects of pure $A^0$ resonance dips in $\ttop$ channel by using the latest search results and the modified NWA.

\section{General formalism }
\label{sec:general}
We study the interference between the continuum and the resonance
in a $2 \to 2$ scattering.
The general expression for the partonic differential cross-section is
\begin{equation}
\label{eq:amp1}
\frac{d{\hat \sigma}}{d z}
= \frac{1}{32 \pi {\hat s}} \sum \left| {\calA}_{\rm bg} e^{i \phi_{\rm bg}}
+ \frac{M^2}{{\hat s}-M^2+iM\Gamma} \cdot {\calA}_{\rm res}e^{i \phi_{\rm res}}
\right|^2\,,
\end{equation}
where $z = \cos\theta^*$,
$\theta^*$ is the scattering angle in the center-of-mass frame,
and we properly sum and average over helicities and colors.
The first term is for the continuum background and
the second is for the resonance with the mass $M$ and the width $\Gamma$.
$\calA_i$ and $\phi_i$ are the magnitude and complex phase of each helicity amplitude, respectively.
We factor out a BW propagator in the resonance amplitude.
The complex phase can be generated
by either loop diagrams or CP-violating interactions.

The interference effects are more clearly shown in the following form:
\begin{eqnarray}
\label{eq:formalizm}
{\hat \sigma} &=&  {\hat\sigma}_{\rm bg} + \frac{M^4}{(\shat-M^2)^2+M^4w^2}  \\ \nn
&&\times\Bigg[  \frac{2(\shat-M^2)}{M^2}\,{\hat \sigma}_{\rm int} c_\phi + {\hat \sigma}_{\rm res}
\bigg(1+\frac{2w}{R}s_\phi\bigg)\Bigg] ,
\end{eqnarray}
where $s_\phi = \sin\phi$, $c_\phi = \cos\phi$, and
\bea
\label{eq:def1}
{\hat \sigma}_{\rm bg, res} &=& \frac{1}{32\pi\shat}
 \int dz \sum \calA_{\rm bg, res}^2,
 \\
\label{eq:phi:definition}
{\hat \sigma}_{\rm int} e^{i\phi} &=&
\frac{1}{32\pi\shat} \int dz \sum \calA_{\rm bg}\calA_{\rm res} e^{i(\phi_{\rm res}
- \phi_{\rm bg})}, \, \\
R &=& \frac{{\hat \sigma}_{\rm res}}{{\hat \sigma}_{\rm int}}~ ,
\quad w\equiv \frac{\Gamma}{M}.
\eea
The $w, R$ and $\phi$ are our key parameters.
If the interference is dominated by a certain helicity amplitude and the resonance amplitude does not depend on the scattering angle $z$ (as is for a scalar resonance),
the $R$ and $\phi$ can be well approximated by the dominant helicity amplitude as
\bea
\label{eq:R:phi:approx}
R\simeq \frac{{ \calA}_{\rm res}}{{ \calA}_{\rm bg}},
\quad \phi \simeq \phi_{\rm res}-\phi_{\rm bg}.
\eea
These imply that $R$ and $\phi$ are the relative strength and phase
between the resonance and continuum, respectively.

Depending on $R$, $\phi$ and $w$ for a given process, various resonance shapes can arise.
The first term in the second line of \Eq{eq:formalizm},
originated from the real-part interference ($c_\phi$), is odd in $\hat{s}$ near $M^2$.
The second term, the sum of the resonance square and the imaginary-part interference ($s_\phi$),
is even in $\sqrt{\hat{s}}$.
The relative strength and sign
between the first and second terms
will determine the overall resonance shape, parameterized well by $R$ and $\phi$.
The simplest example is the well-known purely real-part interference with $s_\phi=0$
(no net relative phase); a peak-dip or dip-peak structure arises and  shifts the
final peak position depending on the size and sign of $\sghat_{\rm int} c_\phi$.

\begin{figure}[t]
\centering
\includegraphics[width=0.46\textwidth]{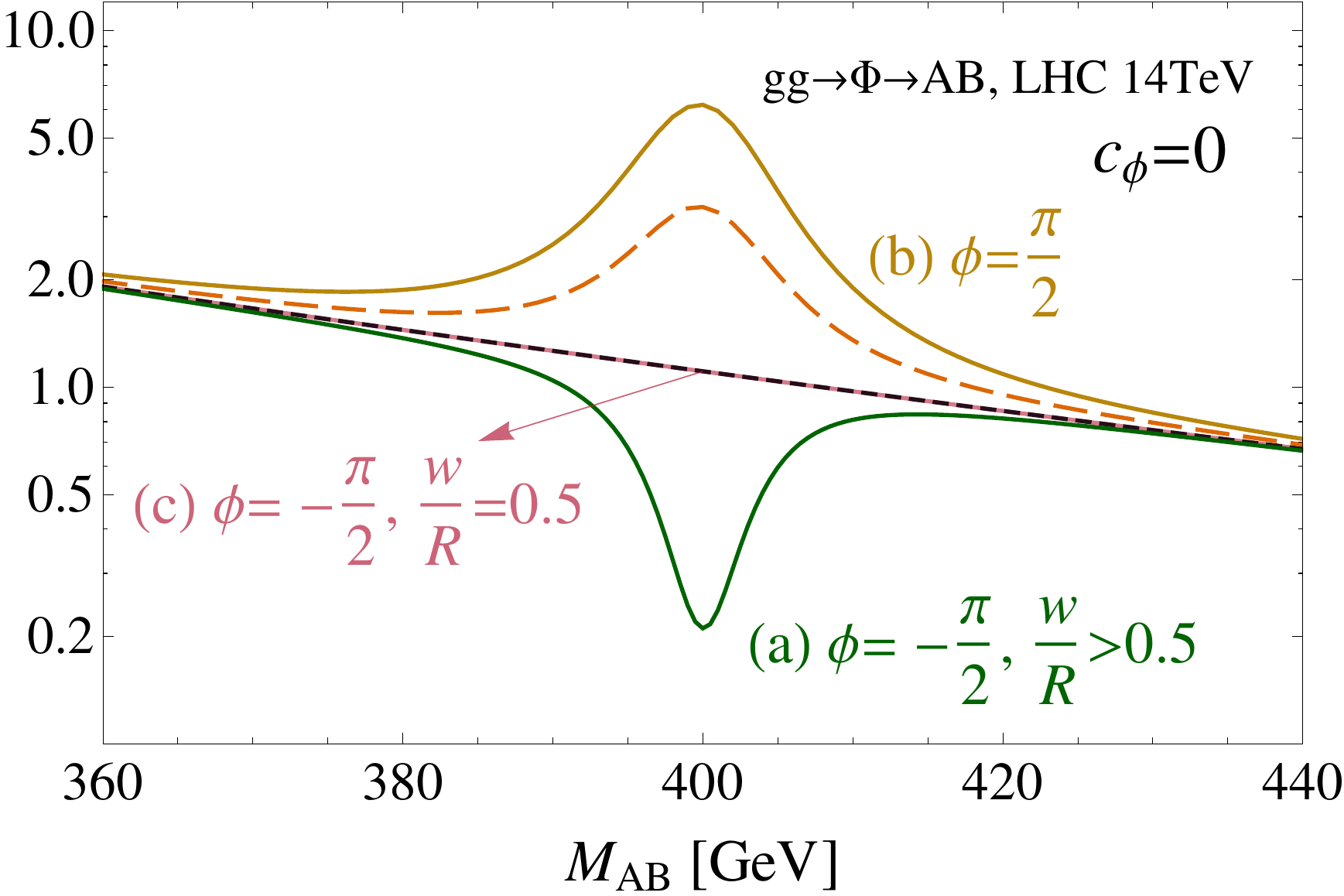}
\vspace*{-0.2cm}
\caption{\label{fig:imaginary}
Resonance shapes from the imaginary-part interferences ($c_\phi=0$)
for $M = 400\gev$, $\Gamma=10\gev$ and $R=0.035$ ($R=0.05$ for the case (c)).
The vertical axis uses arbitrary unit.
Solid lines are the results for (a) a pure dip in \Eq{eq:condition},
(b) a pure enhanced peak ($\phi=\pi/2$), and (c) nothingness ($\phi=-\pi/2,~ w/R=0.5$).
For comparison, we also show a resonance without any interference (orange-dashed)
and the continuum alone (black-dotted).
}
\vspace{-0.2cm}
\end{figure}

On the other hand, the non-zero imaginary interference can produce strikingly different resonance shapes.
First of all, the purely imaginary interference can  produce a \emph{pure resonance dip} if the following conditions are satisfied:
  \beq
\textrm{Pure dip:} \quad \phi = -\frac{\pi}{2}, \quad  \frac{w}{R} > 0.5.
  \label{eq:condition}
  \vspace{-0.1cm}
  \eeq
A pure dip in an example invariant mass distribution is shown in \Fig{fig:imaginary}
by the line (a) for $gg \to \Phi \to AB$ at the LHC 14 TeV. The collider signature would be a deficit rather than an excess in the invariant mass distribution.
The pure dip condition in Eq.~(\ref{eq:condition})
requires the pure negative imaginary interference,
a sizable decay width and relatively small resonance-to-interference strength.

We also show other interesting cases of the pure imaginary interference in \Fig{fig:imaginary}.
As shown by the line (b), the resonance can be a pure peak
if $c_\phi=0$ and  $(1+2w s_\phi/R)>0$ in \Eq{eq:formalizm}. The peak can be enhanced ($\phi = \pi/2$) or suppressed ($\phi = -\pi/2$) compared to the result without interference (orange-dashed). Surprisingly, the resonance can even disappear if $\phi = -\pi/2$ and $w/R=0.5$
denoted by the line (c), in which case both real and imaginary terms vanish in \Eq{eq:formalizm}. These various new features suggest that we change the image of a particle resonance:
its signal does not have to be an excess;
it can be a deficit or even nothing.
We will see in the next section that all of these features can appear
for heavy Higgs bosons.

The NWA is not a good approximation with the non-zero imaginary part of interference. It is because, being an even function in $\hat{s}$ near the resonance, the imaginary part survives under the integration of $\sqrt{\hat{s}}$ across the resonance mass, thus contributing to the total cross-section; on the other hand, the real part practically vanishes under the integration.
We, however, note that the imaginary part adds to the $\hat{\sigma}_{\rm res}$, simply modifying the coefficient of the BW-square by $(1+2 \frac{w}{R} s_\phi)$ in \Eq{eq:formalizm}. Therefore, the factor $(1+2\frac{w}{R} s_\phi)$ serves as the correction factor to the NWA as if the original resonance-square signal were multiplied by the factor. Thus, we suggest the \emph{modified} NWA as:
\beq
 \sigma(ab \to \Phi\to cd)_{\tt w/\,intf} \, = \, \sg(ab\to \Phi)\cdot{\rm Br}(\Phi\to cd) \cdot C,
 \label{eq:mnwa}
\eeq
where the subscript ``{\tt w/\,intf}'' emphasizes that full interference effects are included, and the correction factor $C$ is defined as
\beq
\label{eq:correction}
C \equiv \bigg(1+\frac{2w}{R} s_\phi\bigg).
\eeq
For narrow resonances, the energy dependences of $R, w$ and $\phi$ can be ignored, and we evaluate them at $\hat{s} = M^2$.

The observation of resonance shape can be limited by the resolution of the invariant mass. If the width is much smaller than the experimental resolution, the observable would just be the resonance signal integrated over the experimental bin size. In such case, the correction factor $C$ in \Eq{eq:correction} can also serve as a useful measure of how the resonance would be observed:
\bea
\label{eq:classification}
C < 0 : && \hbox{ deficit}, \\ \nn
C =0  : && \hbox{ nothingness}, \\ \nn
C > 0 : && \hbox{ excess}.
\eea
The ``nothingness'' can be resulted either from the complete disappearance of a resonance ($\phi = -\pi/2, C=0$) or from remaining symmetric dip-peak or peak-dip structures ($\phi \ne - \pi/2, C=0$).
We will use the sign of the $C$ factor to discuss resonance shapes in the general parameter space of heavy Higgs bosons in the next section.

\section{The abnormality of heavy Higgs resonances in the aligned 2HDM}
\label{sec:2HDM:overview}

The interference effects from a resonance particle become significant when
the width is not too narrow.
Heavy neutral Higgs bosons,
which are ubiquitous in many new physics models with the extended Higgs sector,
are good candidates for sizeable total width.
In addition, their production through the gluon fusion at one-loop level
develops a complex phase when the heavy Higgs boson masses are
above the $t\bar{t}$ or $b\bar{b}$ thresholds.
As one of the simplest extensions of the SM,
we consider the heavy Higgs bosons in the CP-conserving aligned Type II 2HDM.

A 2HDM \cite{2HDM} introduces
two complex $SU(2)_L$ Higgs doublet scalar fields,
both of which have nonzero vacuum expectation values $v_1$ and $v_2$.
There are five physical Higgs boson degrees of freedom,
the light CP-even scalar $h^0$,
the heavy CP-even scalar $H^0$, the CP-odd pseudoscalar $A^0$,
and two charged Higgs bosons $H^\pm$.
We consider the resonance shapes of $H^0$ and $A^0$.
The SM Higgs field
is a mixture of $h^0$ and $H^0$ as
\bea
H^{\rm SM} = \sba h^0 + \cba H^0,
\eea
where $\alpha$ is the mixing angle between $h^0$ and $H^0$ and $\tan\beta=v_2/v_1$.
Since the current LHC Higgs data prefers a very SM-like Higgs boson~\cite{2hdm:Higgs:fit},
we simply assume the exact alignment limit, $\sba=1$.
Then we discuss each phenomenology of $H^0$ and $A^0$ in the two-dimensional parameter space of $M_{H/A}$ and $\tan \beta$\footnote{There is another model parameter,
the soft $Z_2$ symmetry breaking term $m_{12}^2$.
The $m_{12}^2$ affects the Higgs triple couplings, which
do not play a significant role in the $H/A$ decays into $\ttop,\bb,\rr$.}.
The Yukawa couplings normalized by the SM ones become
\bea
\label{eq:yukawa}
\yh^H_t = -\frac{1}{\yh^H_b}= -\frac{1}{\tan \beta},
\quad
\yh^A_t = \frac{1}{\yh^A_b}= \frac{1}{\tan\beta
}.
\eea
Note that the $H$-$t$-$\bar{t}$ and $H$-$b$-$\bar{b}$ vertices have the opposite sign
while $A$-$t$-$\bar{t}$ and $A$-$b$-$\bar{b}$ vertices have the same sign.

Important decay modes of $H^0$ and $A^0$ are $\ttop$, $\bb$, $\ttau$,
$\rr$, $\gamma Z$, $hh$ and $Zh$. The $ZZ$ and $WW$ decay modes are forbidden at the tree level in the alignment limit. In the following subsections, we study $\ttop$, $\bb$ and $\rr$ channels to see whether striking resonance shapes are produced and to classify the parameter space as pure dip, nothingness and enhanced peak regions. We also briefly discuss other channels in the last subsection.

\subsection{Dips and nothingness in $gg\to H/A \to t\bar{t}$.}
\label{subsec:tt}
For $M_{H/A} \geq 2 m_t$, the dip signal can arise in $t\bar{t}$ final states~\cite{Gaemers:1984sj,Dicus:1994bm}. We delineate how the dip conditions are satisfied, and we discuss various other resonance shapes in a general parameter space. In the interference between $gg \to H/A \to t\bar{t}$ and the continuum background $gg \to t\bar{t}$, a sizable complex phase is naturally generated from the top quark loop. Since the resonance process is one-loop suppressed compared to the continuum process and the heavy Higgs width is usually ${\cal O}(1)\GeV$, the $R$ is roughly of the same order or smaller than $w$ in most of the parameter space. Therefore, the necessary dip condition $w/R>0.5$ is easily satisfied.

The helicity amplitudes for the one-loop resonance process
$gg \to H/A \to t\bar{t}$
are
\bea
\label{eq:ttsignal}
{\cal M}^{H/A}_{\lm} &=& \delta^{ab} \delta_{ij} \Big(\frac{\alpha_s}{8\pi}\Big)
\yh_t m_t  G_F\sqrt{2 \shat}
\, \widetilde{\cal M}_\lm^{H/A}\\ \nn
&&\quad \times
\frac{\hat{s} }{ \hat{s}-M^2 +i M \Gamma}   \sum_q \hat{y}_q A_{1/2}^{H/A}(\tau_q) ,
\eea
where $\lm$ collectively denotes the helicities of two gluons and two photons, $M=M_{H/A}$,
$\tau_q = M^2/4 m_q^2$, and the loop functions $A_{1/2}^{H/A}$ are given in
Ref.~\cite{Djouadi:2005gj}.
The normalised  nonzero helicity amplitudes  in the chiral representation~\cite{Hagiwara:1985yu} are
\bea
&&~~~~~~~\!\!\!\!\!\widetilde{\cal M}^H_{++++} = -\widetilde{\cal M}^H_{++--} = -\beta_t\,,  \nn \\
\label{eq:ttampH}
&&~~~~~~~\!\!\!\!\!\widetilde{\cal M}^A_{++++} =  \widetilde{\cal M}^A_{++--} = -1\,, \nn \\
&& \widetilde{\cal M}^{H/A}_{----} = -\widetilde{\cal M}^{H/A}_{++++},
~\widetilde{\cal M}^{H/A}_{--++} = -\widetilde{\cal M}^{H/A}_{++--}\,,
\eea
where $\beta_t = \sqrt{1-4m_t^2 /\hat{s}}$.
Obviously, the Higgs contributions have only color-singlet amplitudes.

The helicity amplitudes of the tree level continuum processes are
\bea
\label{eq:tt:bf}
{\cal M}^{\rm bg, \mathbf{1}}_\lm &=& \delta^{ab} \delta_{ij} \frac{4\pi \alpha_s}{2 N_c} \cdot
\widetilde{\cal M}_\lm^{\rm bg},
\eea
where we show only color-singlet components, $\delta^{ab} \delta_{ij}$.
The normalized helicity amplitudes interfering with the Higgs contributions
in Eq.~(\ref{eq:tt:bf}) are
\bea
\label{eq:ttcont}
&&\!\!\!\!\!
\widetilde{\cal M}^{\rm bg}_{++++} = -\widetilde{\cal M}^{\rm bg}_{----} =
\frac{(1+\beta_t) \, m_t \, \hat{s}^{3/2} }{(t-m_t^2) \, (u-m_t^2)}, \\ \nn
&&\!\!\!\!\!
\widetilde{\cal M}^{\rm bg}_{++--} = -\widetilde{\cal M}^{\rm bg}_{--++} =
\frac{(1-\beta_t) \, m_t \, \hat{s}^{3/2} }{(t-m_t^2) \, (u-m_t^2)}.
\eea
The dominant interference is through
$\widetilde{\cal M}_{\pm\pm\pm\pm}$.

\begin{figure}[t] \centering
\includegraphics[width=0.46\textwidth]{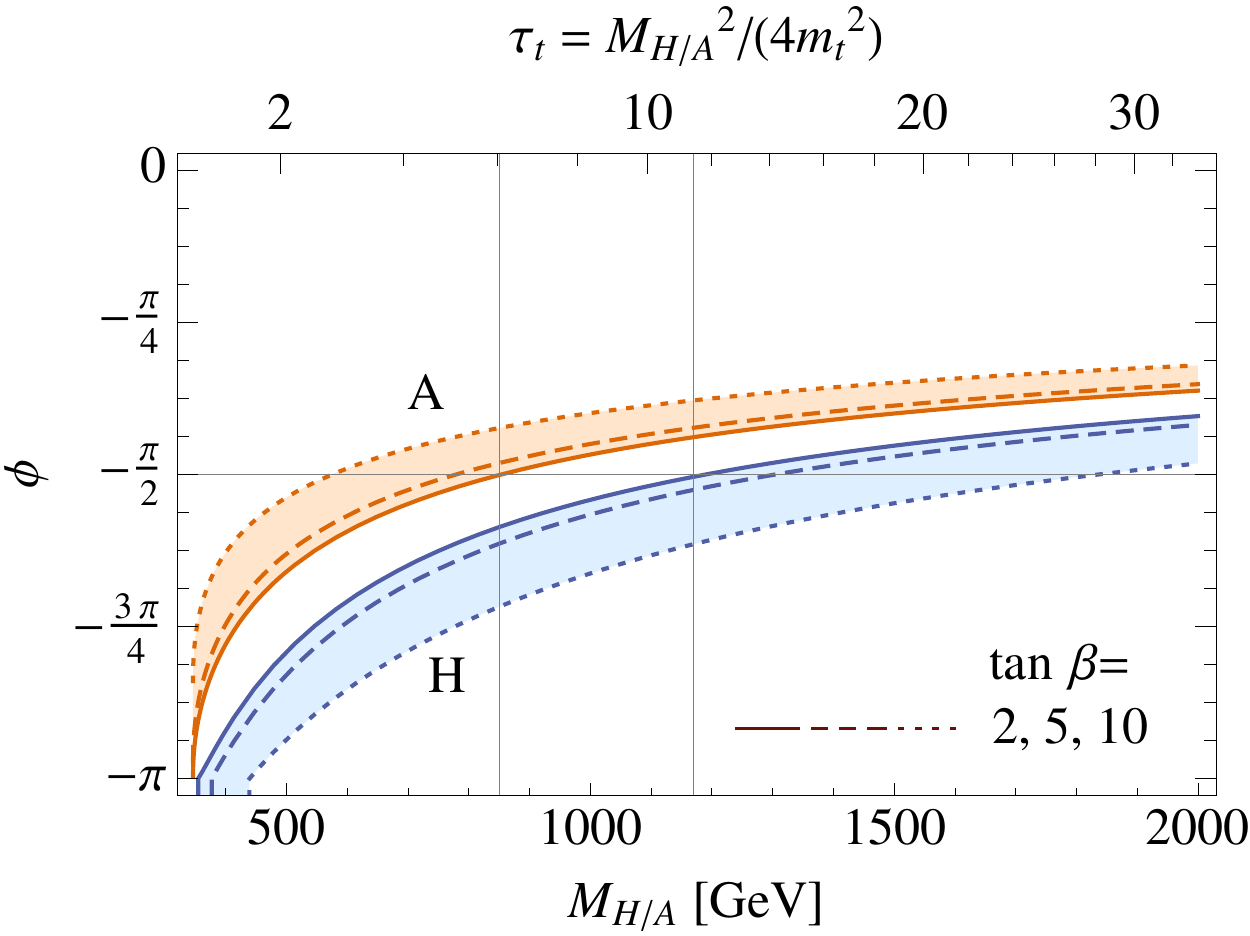}
\vspace{-0.2cm}
\caption{The relative phase, $\phi$ in \Eq{eq:phi:definition}, between the continuum $gg \to t\bar{t}$ and the resonance $gg \to A/H \to t\bar{t}$
as a function $M_{H/A}$
for $\tb=2,5,10$. $M_A=850$ GeV and $M_H=1170$ GeV can achieve $\phi = -\pi/2$ for $t_\beta=2$.  In the upper horizontal axis, we also show the corresponding $\tau_t = M_{H/A}^2/(4m_t^2)$ used in $A_{1/2}^{H/A}(\tau_t)$.
}
\label{fig:phase}\vspace{-0.2cm}
\end{figure}
\begin{figure*}[t]
\centering
\includegraphics[width=0.45\textwidth]{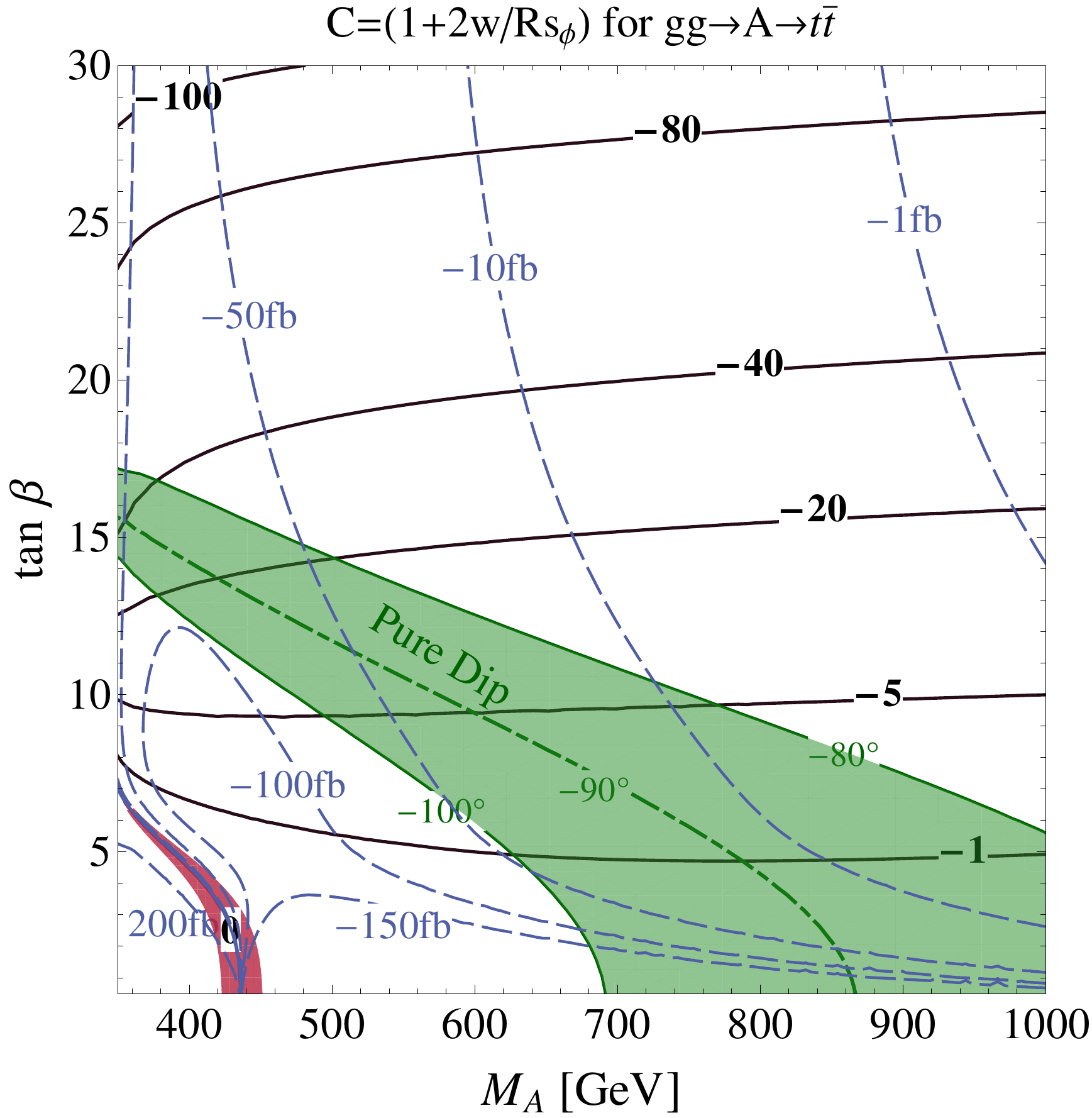} \hspace{0.1cm}
\includegraphics[width=0.43\textwidth]{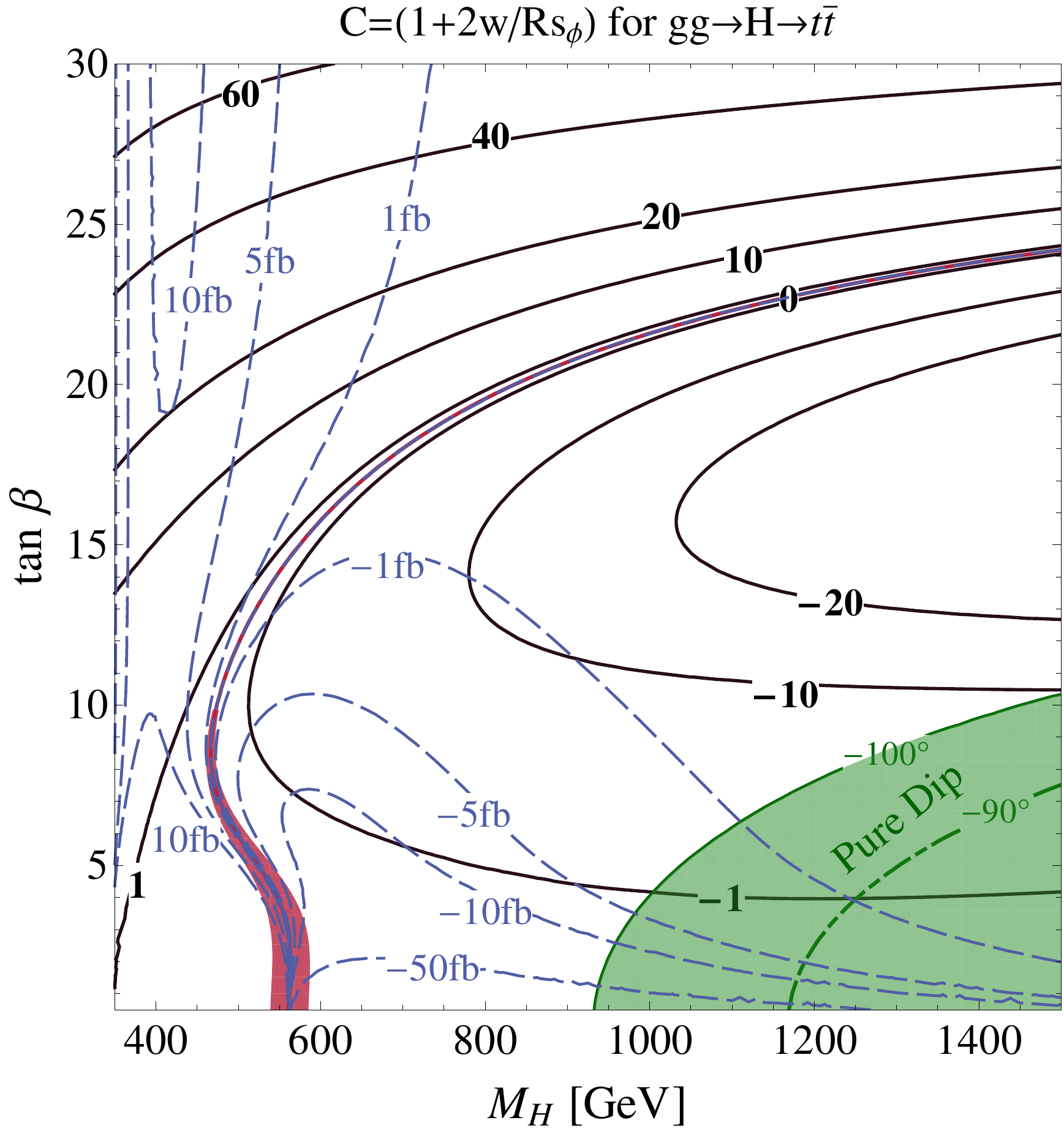}
\vspace*{-0.4cm}
\caption{\label{fig:contour:tt} The $C$ factor in \Eq{eq:correction} (solid) and
the integrated $\ttop$ signal rate (dashed) at the 14 TeV LHC obtained by using the modified NWA in \Eq{eq:mnwa}. They are shown for $A^0$ (left) and $H^0$ (right) Higgs bosons in the aligned Type II 2HDM. Green regions are for nearly pure dips ($-100^\circ \leq \phi \leq -80^\circ$) and red for nearly nothingness ($-0.1 \leq C \leq 0.1$). }\vspace{-0.2cm}
\end{figure*}
\medskip
{\it Pure dips.} We first focus on pure dips and delineate how and where in parameter space the dip conditions are satisfied.
The main sources of the relative phase are (i) the top-quark loop in the resonance process and (ii) the relative minus sign between the resonance and the continuum amplitudes. Interestingly, the dominantly interfering helicity amplitudes for continuum,
${\cal M}_{++++}$ and ${\cal M}_{----}$, have opposite sign to the resonance amplitudes as shown in Eqs.~(\ref{eq:ttampH}) and (\ref{eq:ttcont}).
We plot the resulting relative phase $\phi$ in \Fig{fig:phase}. The relative phase approaches to $-\pi$ for $M_{H/A} \lesssim 2m_t$ with $\tb \lesssim 5$ since the resonance amplitudes are almost real associated with the aforementioned relative minus sign. As the Higgs mass increases, the top-loop imaginary term turns on after $2m_t$ and grows while the real term decreases and eventually flips its sign.
When the real term crosses zero, the condition $\phi = -\pi/2$ is achieved; such solutions are marked as vertical lines for $A^0$ and $H^0$ in \Fig{fig:phase}:
$M_A = 850$ and $M_H=1170$ GeV for $\tb=2$.
In these solutions, the relative minus sign between the resonance and continuum amplitudes is essential.
Obviously, light quark contributions are suppressed by small quark masses, and the tree-level continuum is purely real.

For $\tb \gtrsim 5$, the bottom-quark loop also induces the sizable imaginary part. The bottom-loop phase adds (subtracts) to the top-loop phase for $A^0$ $(H^0)$ because $\hat{y}_b$ has the same (opposite)-sign with $\hat{y}_t$; having already sign-flipped real part, the bottom-loop cancels (adds to) the top-loop real parts.
When the real part of the top loop is canceled by that of the $b$ quark loop,
$\phi = -\pi/2$ is achieved.
  As a result, we see in \Fig{fig:phase} that a somewhat lighter (heavier) $A^0$ $(H^0)$ boson obtains $\phi=-\pi/2$ for $\tan\beta \gtrsim 5$.

\begin{figure*}[t] \centering
\includegraphics[width=0.45\textwidth]{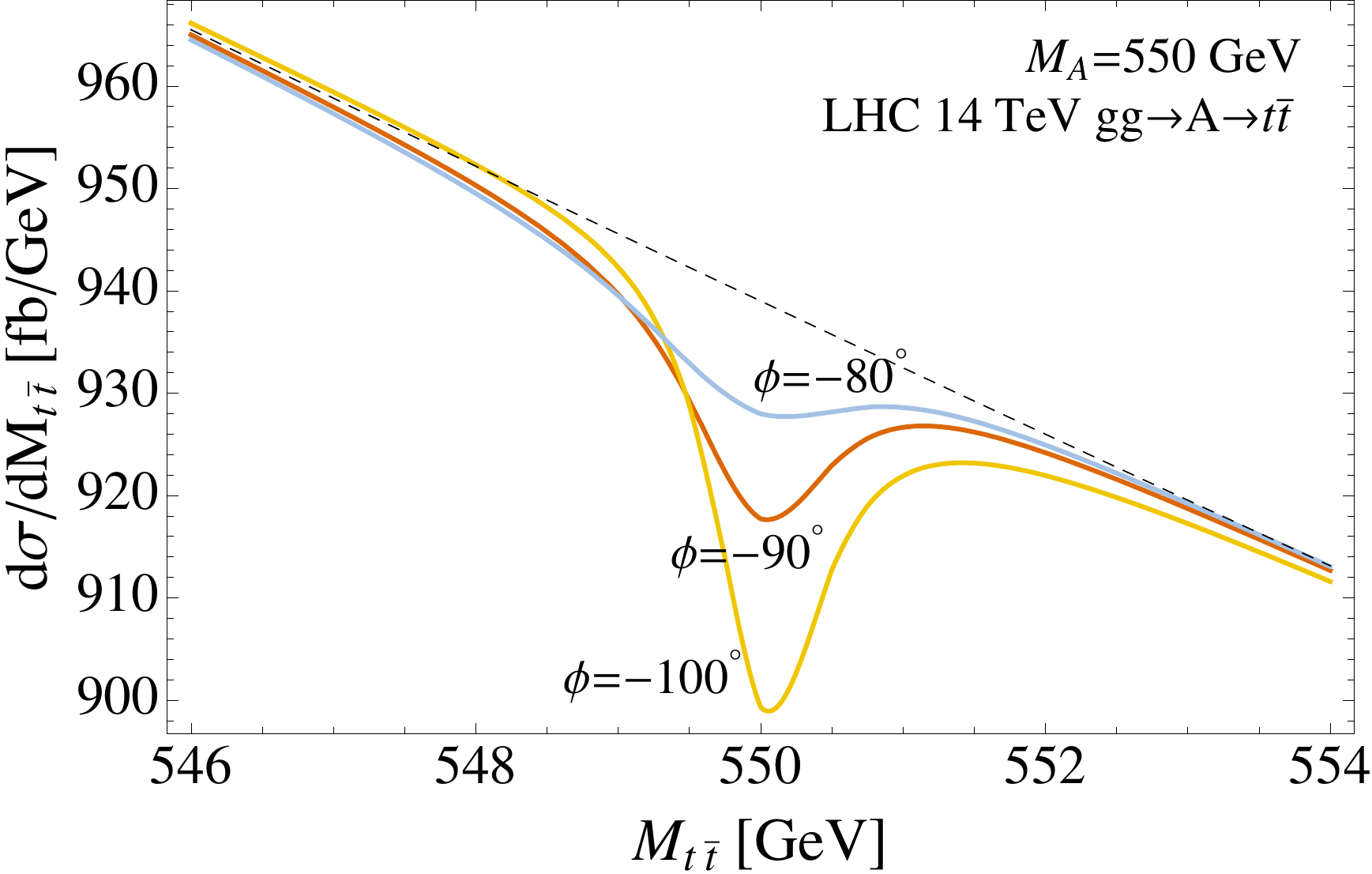}  \hspace{0.1cm}
\includegraphics[width=0.46\textwidth]{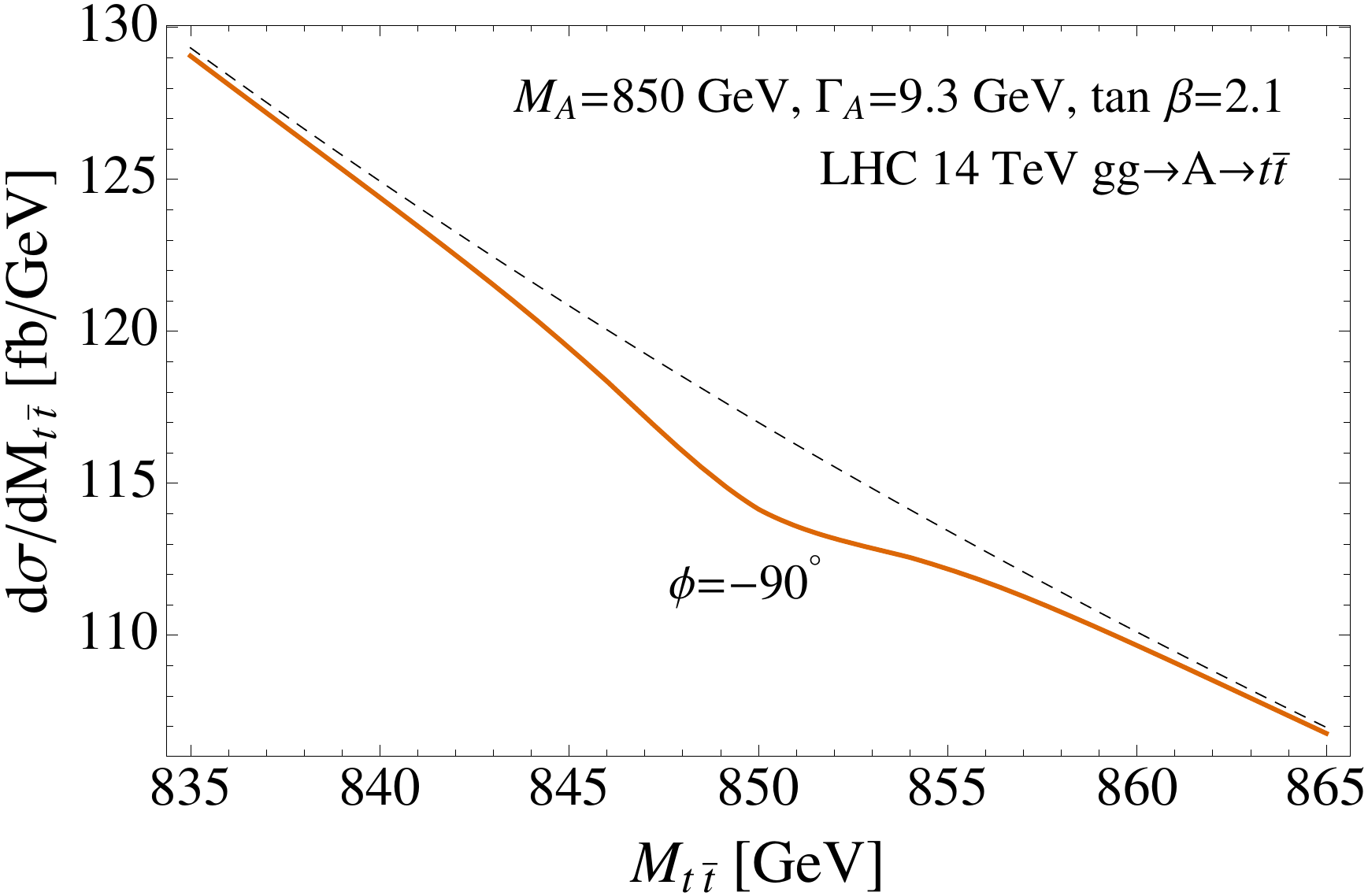}
\vspace{-0.2cm}
\caption{\label{fig:mtg:distribution} Pure dips in the full physical $\ttop$ invariant mass distributions from $pp \to t\bar{t}$
at the 14 TeV LHC; the results include
the resonance $gg\to A \to t\bar{t}$ interfering with the continuum $gg \to t\bar{t}$ as well as the $q\bar{q} \to t\bar{t}$ continuum.
We show for $M_{A}=550\gev$ with $\tb =13.4, 10.6, 7.6$ yielding $\phi= -80^\circ, -90^\circ, -100^\circ$ (left)
and $M_A=850\gev$ with $\tb =2.1$ (right).
The continuum process alone is shown by the dashed lines for comparison. We have used LO results and MSTW PDF and considered no top decays, kinematic cuts and Gaussian smearing.
}
\vspace{-0.2cm}
\end{figure*}

\begin{figure}[t]
\centering
\includegraphics[width=0.46\textwidth]{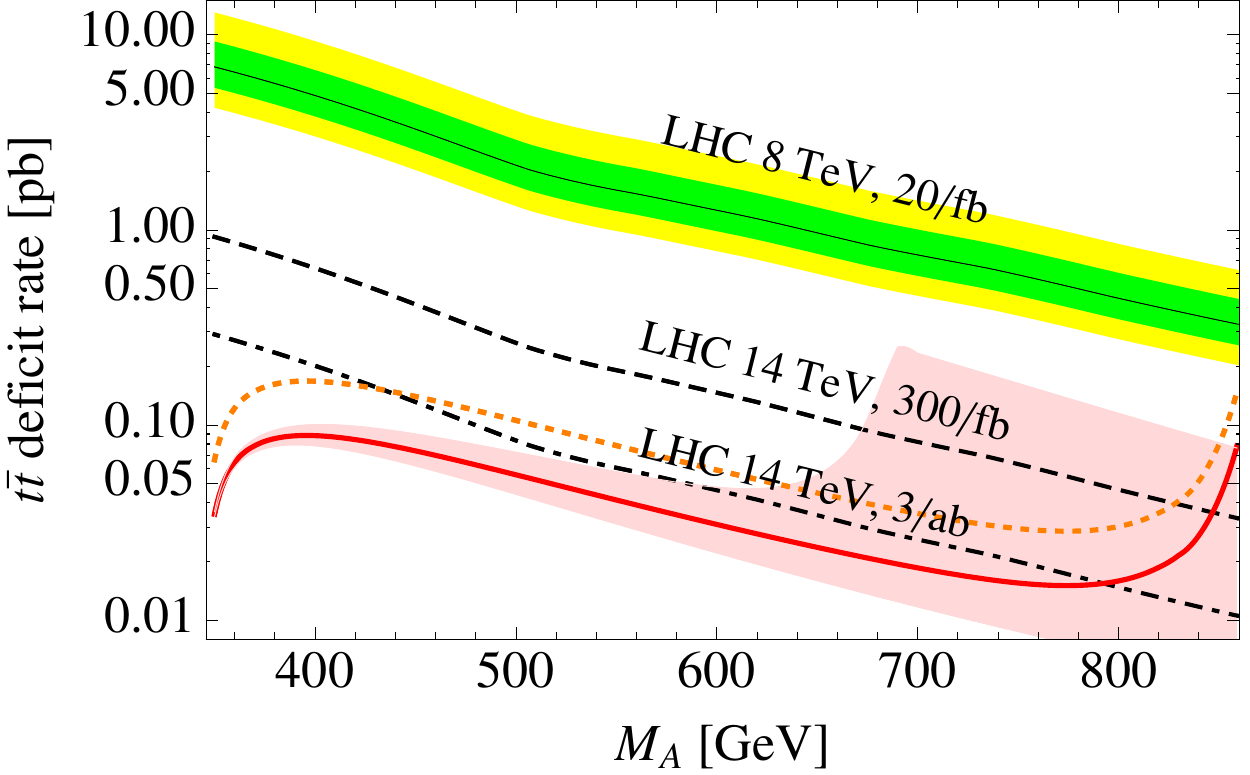}
\vspace*{-0.4cm}
\caption{\label{fig:tt:prospect}
The 95\% CL exclusion prospects of the pure $\ttop$ resonance dip of $A^0$ from the latest 8 TeV 20/fb expected result~\cite{ATLAS:ttbar} and its 14 TeV projections for 300/fb and 3/ab. The modified NWA in \Eq{eq:mnwa} is used to obtain total deficit signal rates at LO (solid red) and with a constant NLO $K=1.9$ (orange dotted). The red band spans $-100^\circ \leq \phi  \leq -80^\circ$ as in \Fig{fig:contour:tt} and is terminated for $\tb \geq 1.5$.
}
\vspace{-0.2cm}
\end{figure}

Figure.~\ref{fig:contour:tt} green bands recap our pure dip discussions so far. The green bands are shown for $-100^\circ \leq \phi \leq -80^\circ$, in which region a resonance will almost look like a pure dip as will be shown in Fig.~\ref{fig:mtg:distribution} left panel. As discussed, any $2m_t \leq M_A \lesssim 850$ GeV can appear as a pure $t\bar{t}$ dip with some larger $\tan\beta$, whereas only the $H$ heavier than about 1170 GeV can appear as a pure dip. The $C$ factor is also shown in \Fig{fig:contour:tt}, and is, of course, negative in the pure dip region. We also show the integrated signal rate using the modified NWA as dashed lines. Along the $\phi=-90^\circ$ contour, the pure dip deficit signal rate decreases with the increasing $M_A$, but the rate increases again near $M_A \gtrsim 800$ GeV. The returning increase is because $\tan \beta$ becomes very small there; the deficit signal rate is also plotted as a red-solid line in \Fig{fig:tt:prospect}. Using the modified NWA rate, we will study pure dip search prospects in the later part of this subsection.


\medskip
{\it Nothingness and other effects.}
 Another interesting resonance shape, nothingness, can also appear in this final state.
 Let us first discuss the $A^0$ nothingness in the left panel of Fig.~\ref{fig:contour:tt}. As $M_A$ becomes lighter or $\tan\beta$ becomes smaller from the pure dip region, the phase deviates from $\phi=-90^\circ$. As $|s_\phi|$ becomes small enough so that the imaginary-part interference becomes comparable to ${\hat \sigma}_{\rm res}$ but still with relative minus sign in \Eq{eq:correction}, then the $C$ factor can vanish. The nothingness region
  spanning $-0.1 \leq C \leq 0.1$ is shown as a red band in the Fig.~\ref{fig:contour:tt} left panel; this region appears to be somewhat parallel to the pure dip region.
 In this region, only the real-part interference remains after the imaginary-part and the resonance-square cancel out ($\phi \ne -\pi/2, C=0$), and the search is likely challenging as the integrated rate is small unless a careful shape analysis can be done with a good experimental resolution. Likewise, the $H^0$ boson in the \Fig{fig:contour:tt} right panel can also show up as nothingness for  $M_H < 550\GeV$ and small $\tan\beta$. This region is also somewhat parallel to the pure dip region.

For both $A^0$ and $H^0$, in the region between nothingness and  pure dip bands, the $C$ factor is negative. Across the nothingness region, the $C$ factor flips its sign, and excesses now will be observed in light enough $M_{A/H} \lesssim 450,~550$ GeV and large enough $\tan\beta$ for $H^0$.

Although $t\bar{t}$ signal estimated from the modified NWA is naively strongest for $\tan \beta \lesssim 5$, the imaginary-part interference introduces potential challenges; complicated resonance shapes and possibly small integrated rates near the nothingness region will need careful shape analysis -- the pure dip can be an interesting exception as will be discussed.
The experimental search in these regions may be aided by other production and/or decay channels that are not much complicated by interference effects; see e.g. Refs.~\cite{Craig:2015jba,Hajer:2015gka} for the study of $t\bar{t}H/A \to t \bar{t} t \bar{t}$ and $b\bar{b} H/A \to b\bar{b} t \bar{t}$ associated productions.

\medskip
{\it Pure dip demonstration.}
We now demonstrate the pure dips in the physical $M_{\ttop}$ distributions in \Fig{fig:mtg:distribution}
for the $A^0$ boson.
Results are for $M_{A}=550\gev$ and $M_A = 850$ GeV. For $M_A=550$ GeV shown in the left panel, we show three results for $\phi=-80^\circ, -90^\circ, -100^\circ$ corresponding to $\tan\beta =13.4,\, 10.6,\, 7.6$.
For $M_A=850\gev$, we used $\tan\beta =2.1$ for $\phi=-90^\circ$. Indeed, very clear dips are produced. The left panel also demonstrates that any resonance with $-100^\circ \leq \phi \leq -80^\circ$ appear almost like pure dips.
The integrated deficit rates for those three $550\GeV$ shapes are $-27\fb$, $-37\fb$, $-53\fb$,  predominantly determined by $\tan\beta$, not by $\phi$.
The smaller $\phi$ (such as $\phi \sim -100^\circ$), the smaller $\tan\beta$ is needed and the larger signal rates are obtained. The $850 \, \GeV$ shape has a similar integrated deficit rate $-37\fb$. Although its shape is broader than the $M_A=550\GeV$ case,
its discovery prospect is higher due to smaller background as will be discussed. In the same  \Fig{fig:mtg:distribution}, we also show the whole $pp \to \ttbar$ continuum result by dashed lines for comparison.  The leading order (LO) results without top decays, kinematic cuts on the top and any Gaussian smearing are used. We have adopted MSTW LO PDF with scales $\mu_{R,F} = M_A$.

\medskip
{\it Pure dip LHC prospects.}
In general, the collider study of heavy Higgs $\ttop$ resonances is complicated by the complex resonance shapes and the large smearing of the top pair invariant mass distribution. The latest experimental resolution of the invariant mass is about 6--8\%~\cite{ATLAS:ttbar,Chatrchyan:2013lca}, which is bigger than any Higgs widths in the parameter space shown in \Fig{fig:contour:tt} ($w \lesssim 1\%$). Thus, delicate structures of resonances are smeared out and signals are distributed over a few bins. The modified NWA rate is not simply the observable for general resonance shapes.
We refer to Refs.~\cite{Bernreuther:1997gs,Craig:2015jba} for earlier preliminary studies on this channel.

Nevertheless, pure resonance dips can be well studied with the modified NWA, based on the latest resonance searches~\cite{ATLAS:ttbar,Chatrchyan:2013lca} that do \emph{not} take into account any interference effects.  The current searches model a scalar resonance as a pure BW peak without any interference effects, but they are inherently sensitive to deficits too~\cite{ATLAS:ttbar,Chatrchyan:2013lca}.
A pure dip has the pure BW shape with a negative sign as if no interferences existed. Thus, presumably, the current search techniques and results without interferences can be applied to pure peaks and dips equally well.

The current searches then take into account all realistic effects from detector resolution, smearing and selection cuts, and provide upper limits on the true signal rates of pure peaks (and dips) before any selection cuts. We can use the modified NWA in \Eq{eq:mnwa} to obtain pure dip signal rates and simply compare them with the reported experimental upper bound.

The result is shown in \Fig{fig:tt:prospect}. The \emph{expected} result of the latest 8 TeV resonance search~\cite{ATLAS:ttbar} is shown with its error bands. The integrated pure dip signal rates are shown as the thick red line. As discussed, the rate decreases with the $M_A$, but it rapidly increases back after $M_A \gtrsim 800$ GeV, where the $\tan\beta$ satisfying the pure dip conditions becomes very small. We also show a red band around the pure dip line for $-100^\circ \leq \phi \leq -80^\circ$; the upper (lower) red band boundary is from $\phi = -100^\circ(-80^\circ)$. The red band is terminated to have $\tan\beta \geq 1.5$. We have used LO results for signals. But to roughly grasp the potentially important next-to-leading-order (NLO) effects, we also show the signal rates by multiplying an assumed constant NLO $K=1.9$ factor~\cite{Djouadi:2005gj,Song:2014lua}
by the orange-dotted line. It is clear that the latest results do not constrain pure dips yet.

We project the latest 8 TeV expected result~\cite{ATLAS:ttbar} to 14 TeV high-luminosity stages. In the projection, we assume that errors are dominated by statistics and that the $\ttop$ resonance search is characterized most importantly by a single resonance mass scale $M$ so that both signal-to-background ratio and cut efficiencies remain more or less constant over a wide range of proton-proton collision energy; see discussions and validations in Refs.~\cite{Jung:2013zya,colliderreach}. Simply speaking, we use the following relation to obtain the projected upper bound $\sigma_{\rm bound}^{i}$ (superscripped by $i$) from the current (superscripped by $j$) upper bound $\sigma_{\rm bound}^j$ as
\beq
\frac{\sigma_{\rm bound}^{i}}{\sigma_{\rm bound}^{j}} = \sqrt{ \frac{ {\cal P}_{gg}^{i} }{ {\cal P}_{gg}^{j} } } \sqrt{ \frac{{\cal L}^{j}}{ {\cal L}^{i} } },
\label{eq:scaling}
\eeq
where the data luminosity ${\cal L}^k$ and the gluon-gluon parton luminosity function ($k=i,j$)
\beq
{\cal P}^k_{gg} \, \equiv \, \int_{\tau^k}^1 dx \, \frac{\tau^k}{x} f_g(x, Q^2) f_g(\tau^k/x,Q^2)\,,
\label{eq:Pab}
\eeq
for $\tau^k \, \equiv \, M_{H/A}^2 /s^k $.
The 14 TeV projections with $300\,{\rm fb}^{-1}$ and $3\,{\rm ab}^{-1}$ are shown in \Fig{fig:tt:prospect}. The projections  imply that $\sim {\cal O}(100-1000)$/fb of data at the 14 TeV LHC can probe the large part of the pure dip region, especially for $M_A \gtrsim 600$ GeV.  So far, we have ignored any kinematic differences between the top quark produced from resonance-square and that from the imaginary-part interference. We will present a detailed study in our future publication~\cite{our:upcoming}.

Although the pure dip prospect in \Fig{fig:tt:prospect} can be affected by various others such as systematics and kinematics, we think it is a good guideline for future searches and the method used for this study is a remarkable application of our results.

\subsection{Dip, nothingness, or peak  in $gg\to H/A \to \bb,\rr$}
\label{subsec:bb:rr}

\begin{figure}[t]
\centering
\includegraphics[width=0.46\textwidth]{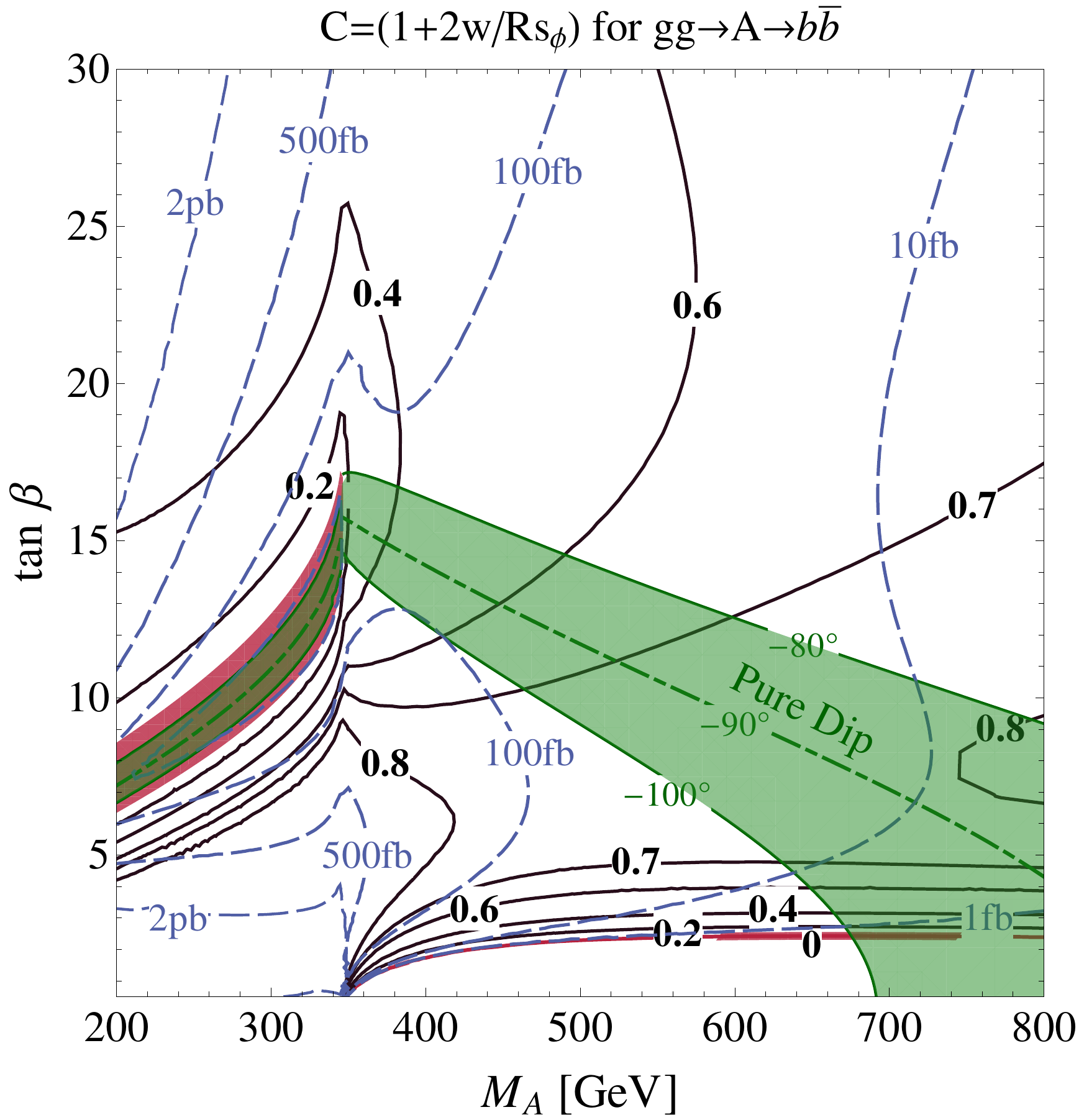}
\vspace*{-0.4cm}
\caption{\label{fig:contour:bb:A} The $C$ factor in \Eq{eq:correction} (solid) and
the integrated $\bb$ signal rate (dashed) at the 14 TeV LHC obtained using the modified NWA in \Eq{eq:mnwa}. Green region is for nearly pure dips $(-100^\circ \leq \phi \leq -80^\circ)$
and red for nearly nothingness $(-0.1\leq C \leq 0.1)$.
}\vspace{-0.2cm}
\end{figure}

The $gg\to H/A \to \bb$ process through the triangle diagram at one loop level
interferes with the $gg \to \bb$ background at tree level.
The helicity amplitudes of the signal and the background
are the same as Eqs.~(\ref{eq:ttsignal})$-$(\ref{eq:ttcont})
with the replacement of
$m_t\to m_b$, $\yh_t\to \yh_b$, and $\beta_t \to \beta_b$.
In Fig.~\ref{fig:contour:bb:A}, we show the $C$ factor and the integrated
signal rate for $A^0$ using the modified NWA.
In the whole parameter region of $200\gev < M_A < 800\gev$,
the $C$ factor is always small positive, like $0.2\sim 0.8$.
We have the smaller peak than the BW one.
Above the $\ttop$ threshold,
the small branching ratio of $A^0 \to \bb$ itself
suppresses the modified signal rate further.
For $M_A>1\tev$, the signal rate is very small to be below 1 fb.

If $M_A < 2 m_t$,
there is a nearly nothingness region at $\tb\simeq 10$ having $0.01<C<0.1$. 
In this region $\phi\simeq \pi/2$, thus almost a complete destruction occurs.
This can be understood by the loop function.
 $A_{1/2}^{A}(\tau_t)$ is a real positive number of the order of one for $M_A  < 2 m_t$.
$A_{1/2}^{A}(\tau_b)$ has
negative real part and positive imaginary part, both being of the order of $0.01$.
Since $\yh_t$ and $\yh_b$ have the same sign,
$\tan\beta \simeq 10$ brings about the cancellation of the real-part interference.
Then, we have a pure negative imaginary interference $(\phi=-\pi/2)$ and sufficiently small $R$ value to fulfil the remaining nothingness condition $(w/R=0.5)$.

For $H^0$ with opposite sign $\yh_t$ and $\yh_b$,
the correction factor $C$ is alway positive in the range of $0.6\sim 1.5$.
Minor peaks will appear.
The current $\bb$ resonance search
for $M_{bb} > 1\tev$ puts the new physics bound on $\sg \times \br \times \mathcal{A}$
about a few hundred fb, where $\mathcal{A}$ is the acceptance.
The small $C$ factor worsens the poor observability.
However, the $\rr$ collider,
which is expected to have much higher sensitivity,
can probe the process $\rr \to H/A \to\bb$ interfering from
the continuum $\rr\to\bb$.
This has very similar features to $gg \to H/A \to\bb$ except for the color factor.

\begin{figure*}[t] \centering
\includegraphics[width=0.45\textwidth]{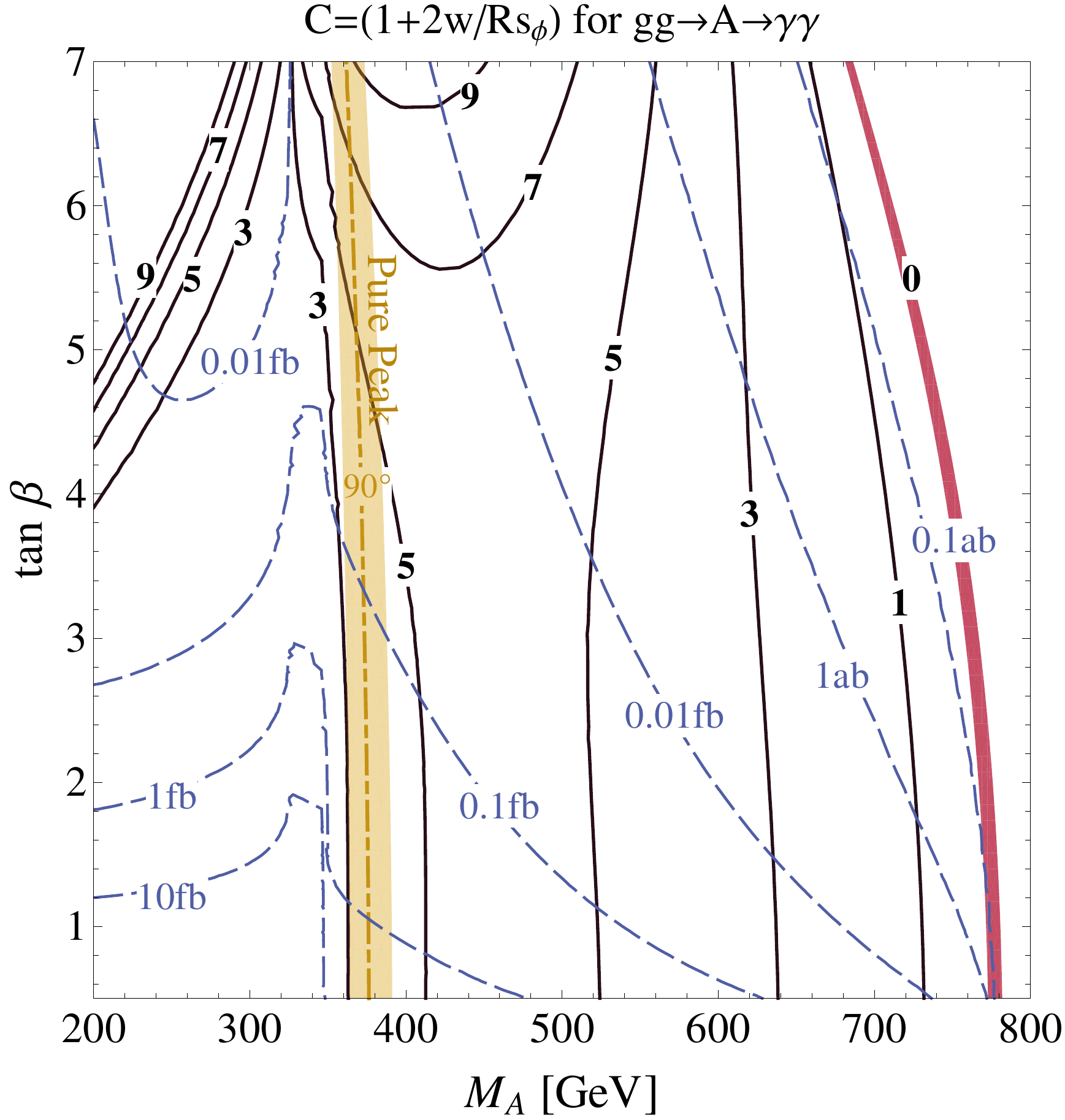}  \hspace{0.1cm}
\includegraphics[width=0.45\textwidth]{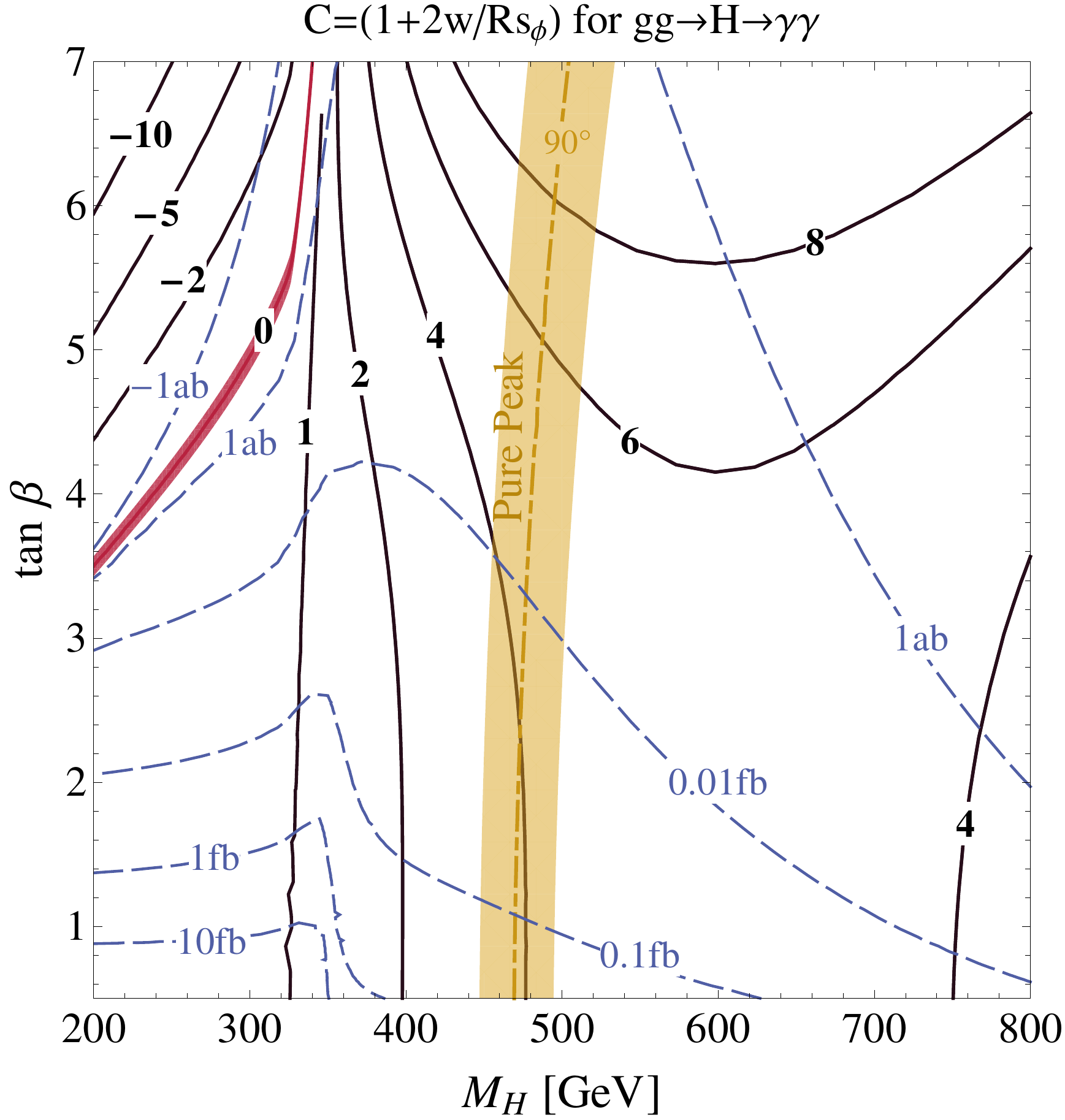}
\vspace{-0.2cm}
\caption{\label{fig:contour:rr}  The $C$ factor in \Eq{eq:correction} (solid) and
the integrated $\rr$ signal rate (dashed) obtained by using the modified NWA
for the $A^0$ (left) and $H^0$ (right) bosons in the aligned Type II 2HDM at the 14 TeV LHC.
Yellow regions are for nearly pure peaks $(80^\circ \leq \phi \leq 100^\circ)$
and red for nearly nothingness $(-0.1\leq C \leq 0.1)$.
}
\vspace{-0.2cm}
\end{figure*}

Another potentially important decay channel of the heavy $H^0$ and $A^0$ is
into $\rr$.
The resonance process is at the two-loop level
and the continuum $gg \to \rr$ is through the box-type diagram at one loop level.
Their leading order helicity amplitudes are given
in Refs.~\cite{Dicus:1987fk,Martin:2012xc}.
We find that ${\cal M}^{H/A}_{\pm \pm \pm \pm}$ give the dominant interference,
which have positive sign relative to the background.

Figure \ref{fig:contour:rr} shows the correction factor $C$ and
the modified single rate $\sg( pp \to  H/A \to \rr)$ by
the interference effects at the 14 TeV LHC.
Here we consider $0.5 < \tb < 7$
since the large $\tb$ severely suppresses both production and decay so that the signal rate becomes very tiny. 

Above the $\ttop$ threshold,
the most parameter space has positive values of $C$ for $A^0$ ($H^0$) if $M_{A^0} <700\gev$ ($M_{H^0}<1000\gev$).
In the $pp \to  H/A \to \rr$ with
two triangle diagrams,
there are three kinds of contributions,
top-triangle times top-triangle, top-triangle times $b$-triangle, and
$b$-triangle times $b$-triangle.
When the top-contribution is dominant,
the total complex phase $\phi$ is just
twice the phase from a single loop function $A^{H/A}_{1/2}$.
Before $\Re e \, A^{H/A}_{1/2}(\tau_t)$ crosses zero,
the argument of $A^{H/A}_{1/2}(\tau_t)$ is in $[0,~\pi/2]$.
Since the relative sign with the background is also plus,
the resulting $\phi$ leads to positive $C$ factor.

Below the $\ttop$ threshold,
the $C$ factor for $A^0$ stays positive.
Although the $C$ factor can be as large as ten at $\tb\simeq 7$,
the signal rate is too small below 0.01 fb,
which is practically impossible to probe.
For $\tb=0.5\sim 3$, the $C$ factor is almost constant, about one:
we do not have higher peak structures than the BW one.
For $H^0$,
on the contrary,
there is a nothingness strip for $\tb =4\sim 6$
where the $b$ quark contribution to the loop function becomes sizable
to cancel the top quark contribution.
Above the nothingness line,
there is a region for dip-like resonances.
Very the small modified NWA rate makes it almost impossible to probe them.

\subsection{Other candidate processes for heavy Higgs dips} \label{sec:higgschannels}

We have so far explicitly studied $\ttop$, $b\bar{b}$ and $\rr$ channels in the alignment limit of 2HDM. In this subsection, we briefly discuss what other hadron collider processes and what modifications of the aligned 2HDM model can produce Higgs resonance dips. The discussion will also show how the dip conditions derived from our general formalism can be usefully applied to a wide range of processes and models.

The $gg \to H \to WW, \, ZZ$ with the heavy SM-like Higgs boson were found to produce peak like resonances~\cite{Campbell:2011cu,Kauer:2015hia,Glover:1988rg}. The immediate reason is $R\sim 1$ -- both resonance signal and continuum are one-loop processes; thus, only a very broad resonance with $w \sim 1$ can satisfy the dip condition $w/R > 0.5$ in \Eq{eq:condition}. But the $R$ can be suppressed if the heavy Higgs boson, $H^0$ in 2HDM, can have non-zero but suppressed $WW, ZZ$ couplings proportional to $\cos(\beta - \alpha)$ (deviating from the alignment limit). The relative phase arises from the quark loops in the $gg \to H$, the sign of $\cos(\beta - \alpha)$ and the unknown relative sign between the resonance and the continuum amplitudes. The loop phase from $gg \to H$ can be $\pm 90^\circ$ as discussed. As one can somewhat freely choose the sign of $\cos(\beta-\alpha)$, the unknown overall sign can perhaps be chosen to satisfy the dip condition $\phi = -90^\circ$. Therefore, pure $ZZ,WW$ dips may be produced in this kind of models, but a dedicated calculation is needed to check this.

The $gg \to H \to \rr$ for the heavy Higgs in the 2HDM with $M_H \sim$ 200 - 300 GeV can also produce pure resonance dips if we deviate from the alignment limit (\Fig{fig:contour:rr} shows that it is not in the alignment limit). Away from the alignment limit, the $H \to \rr$ decay one-loop diagram can have both $W$- and top-loops. The main phase arises from the $W$-loop, but the real parts can be canceled between the $W$- and top-loops, depending on $\cos(\beta-\alpha)$ and $\tan \beta$. The $R$ is already loop-suppressed as desired; thus, pure dip conditions can be satisfied. We have numerically checked that this is indeed possible. Specifically, $M_H =200$ GeV, $\cos(\beta - \alpha) \simeq -0.15$ and $\tan \beta \simeq 1.4$ can produce the pure Higgs dip; and the dip conditions can be satisfied for any $-0.3 \lesssim \cos(\beta - \alpha) \lesssim 0$ and corresponding $1 \lesssim \tan \beta \lesssim 10$ for $M_H \sim$ 200 - 300 GeV. The collider search is, however, likely challenging in the near future because the signal rate is small, limited by the Higgs coupling constraints on the size of $|\cos(\beta - \alpha)| \lesssim 0.15$~\cite{2hdm:Higgs:fit} and by partial cancelation of real parts.

The $gg \to H \to Z\gamma$ is similar to the $\rr$ channel. The resonance is a two-loop process while the continuum is an one-loop process, giving a loop-suppressed $R$ as desired. As soon as the overall sign comes out correctly, Higgs dips can perhaps be produced. But a dedicated study is needed to check this.

Other processes with the light Higgs boson in the final states such as $gg \to H \to hh$ and $gg \to A \to Zh$ may have similar conclusion as the $ZZ$ and $WW$ channels. Both the resonances and continuum are one-loop processes, giving $R \sim 1$; but the $\cos(\beta-\alpha)$ suppression may play a proper role in producing a dip. But the reconstuction of heavy Higgs resonances in these channels might be more challenging than in the $ZZ$ channel. 

Another important Higgs channel is $gg \to H/A \to \tau \tau$. 
Since the interfering continuum does not exist up to the one-loop order, the interference is small and $R\gg1$. Thus, Higgs bosons will always show up as pure BW peaks, and the ordinary NWA will be working well.

Lastly, various associated productions, such as $t\bar{t} H$, are tree-level processes and may not have large enough complex phases. Without additional CP violations, the imaginary interference effects will be absent.

\section{Conclusions}
\label{sec:conclusions}

We have identified new resonance shapes -- pure dips, nothingness and enhanced pure peaks -- arising from the resonance-continuum interference with a relative phase and derived their production conditions.
We have demonstrated that all of these new shapes can generally show up from heavy Higgs bosons, $A^0$ and $H^0$, in $\ttop$, $\bb$ and $\rr$ channels. The pure dip conditions are also applied to various other Higgs processes for the discussion of whether Higgs dips can be produced. 

In addition, we have modified the NWA to work with the non-zero imaginary part of interference in \Eq{eq:mnwa}; it is now possible to easily estimate the total signal rate of the resonance including the imaginary-part interference. The multiplicative correction factor $C$, defined in \Eq{eq:correction}, could also categorize the resulting collider observables as excesses, deficits and nothingness. 

The collider searches of dip-like heavy Higgs resonances are most promising in $\ttop$ channel but, in general, need dedicated shape analyses. Remarkably, for pure $\ttop$ dips of $A^0$ bosons, it was useful to apply the modified NWA and the latest $\ttop$ resonance searches (that do not even account for any interference effects yet) to obtain LHC constraints and prospects (\Fig{fig:tt:prospect}). It was possible because a pure dip has the negative BW shape as if no interferences existed. We hope that this pure dip study in $\ttop$ can be seriously considered by LHC experiments and shed more light on general studies of complex $\ttop$ Higgs resonances.

A resonance dip or nothingness, a special result of the imaginary-part interference,
is an overlooked phenomenon in heavy (Higgs) resonance searches. The imaginary-part interferences will be more important in new physics models;
presumably, heavier resonances from new physics beyond the SM are broader
and have more sources of imaginary terms from loop corrections
and new CP violations.
Combined with other measurements,
resonance shapes will add valuable information on new physics.

\acknowledgements
SJ is supported by the National Research Foundation of Korea under grant 2013R1A1A2058449,
and YWY and JS are supported by NRF-2013R1A1A2061331 and in part by NRF-2012R1A2A1A01006053 (YWY). We thank KIAS Center for Advanced Computation for providing computing resources.


\end{document}